\documentclass[aps,pra,showpacs,twocolumn,amsmath,amssymb,superscriptaddress,footinbib]{revtex4}
\usepackage[dvips]{graphicx}
\usepackage[usenames]{color}
\usepackage{amssymb}
\usepackage{amsmath}

\def\be{\begin{equation}}
\def\ee{\end{equation}}
\def\bea{\begin{eqnarray}}
\def\eea{\end{eqnarray}}
\def\bi{\begin{itemize}}
\def\ei{\end{itemize}}
\def\bin{\begin{enumerate}}
\def\ein{\end{enumerate}}
\def\be{\begin{equation}}
\def\ee{\end{equation}}
\def\bea{\begin{eqnarray}}
\def\eea{\end{eqnarray}}
\def\bi{\begin{itemize}}
\def\ei{\end{itemize}}
\def\bin{\begin{enumerate}}
\def\ein{\end{enumerate}}
\def\la{\langle}
\def\ra{\rangle}

\begin{document}
\title{Huge quantum particle number fluctuations in a two-component Bose gas in a double-well potential}
\author{Pawe\l{} Zi\'n}

\affiliation{Soltan Institute for Nuclear Studies, Ho\.za 69,
00-681 Warsaw, Poland}

\author{Bart\l{}omiej Ole\'s}
\affiliation{
Instytut Fizyki imienia Mariana Smoluchowskiego and
Mark Kac Complex Systems Research Center,
Uniwersytet Jagiello\'nski, ulica Reymonta 4, PL-30-059 Krak\'ow, Poland}

\author{Krzysztof Sacha}
\affiliation{
Instytut Fizyki imienia Mariana Smoluchowskiego and
Mark Kac Complex Systems Research Center,
Uniwersytet Jagiello\'nski, ulica Reymonta 4, PL-30-059 Krak\'ow, Poland}

\date{\today}

\begin{abstract}
Two component Bose gas in a double well potential with repulsive interactions may undergo
a phase separation transition if the inter-species interactions outweigh the
intra-species ones. We analyze the transition in the strong interaction limit within
the two-mode approximation.
Numbers of particles in each potential well are equal and constant. However, 
at the transition point, the ground state of the system reveals huge fluctuations of 
numbers of particles belonging to the different gas components. That is,
probability for observation of any mixture of particles in each potential well becomes uniform.
\end{abstract}


\maketitle


\section{Introduction}\label{s1}

Ultra-cold dilute gases of bosonic atoms constitute perfect
systems for experimental and theoretical investigations of various phenomena of quantum many body problems \cite{leggett01}.
From the viewpoint of quantum computing and interferometry an especially
relevant subject is quantum fluctuations \cite{ladd10, dunningham04}.

Most experimental studies of fluctuations concentrated on systems of cold atoms in double well \cite{oberthaler07} and optical lattice potentials \cite{bloch08}.
In the former system squeezed states were predicted and produced, with particle
number fluctuations (i.e. uncertainties of populations of the potential wells)
turning from poissonian to sub-poissonian
\cite{esteve08, orzel01, choi05}.
The latter system reveals a superfluid to Mott insulator transition \cite{SFMI, greiner02, gerbier06}
with enhanced phase fluctuations but with decreasing particle number fluctuations.

In the present paper we focus on a system where the total particle number is fixed but
occupation of certain single particle states reveals considerable quantum fluctuations.
We are interested in a system where the mean field theory predicts symmetry breaking \cite{break, walls98, gordon99}
and the symmetry broken solutions are degenerated and form a Hilbert subspace 
parameterized by a continuous parameter. 
If the occupation of single particle states varies a lot as we move in the degenerate subspace
than huge particle number fluctuations can be expected in the exact quantum many body eigenstates. 
Degenerate subspace parameterized by a continuous parameter appears in spin-1 Bose gas with an 
anti-ferromagnetic interaction \cite{castin01} or in scalar condensates with solitonic solutions \cite{castin01,dziarmaga10, miszmasz}. 
Attractive single component Bose gas in a symmetric double well potential reveals also huge particle number fluctuations
but it constitutes a slightly different example \cite{ho04, jack05}. There the degeneracy is small, i.e. the degenerate subspace 
is two dimensional, and the particle number fluctuations correspond to random localization of all particles in one of 
the potential wells in different experimental realizations.
In all these examples the correct mean field theory reduces to the Gross-Pitaevskii equations \cite{leggett01}. 
In the present paper we consider a Bose gas system where the Gross-Pitaevskii equation is not a correct mean-field description,
that is, a two-component Bose gas in a double-well potential in the strong interaction limit.

In Sec.~\ref{model} we present a theoretical model for a two-component Bose gas in a double-well potential. 
In Sec.~\ref{EH} we derive the effective Hamiltonian using second order perturbation theory valid in the strong interaction limit. 
In Sec.~\ref{CL} we analyze its mean field (classical) limit and identify phase transition region.
It turns out that the mean-field solutions reveal continuous degeneracy at the transition point.
We deduce the exact ground state of the system in Sec.~\ref{QGS} and show that the particle number fluctuations are indeed huge at the critical point. 
In Sec.~\ref{V} we estimate the range of parameters where the predicted fluctuations can be observed and in Sec.~\ref{Con} 
the results presented in the paper are summarized.

\section{The model}\label{model}

The Hamiltonian of a two component Bose gas in a symmetric double well potential, 
in the tight binding approximation, takes the form of the Bose-Hubbard model
\begin{eqnarray}
\hat H &=& - \ \frac{J}{2} \left(\hat a_1^\dagger\hat a_2+
\hat a_2^\dagger\hat a_1 + \hat b_1^\dagger\hat b_2+
\hat b_2^\dagger\hat b_1\right)
\cr &&
+\frac{U}{2}\left( \hat a_1^\dagger \hat a_1^\dagger \hat a_1 \hat a_1 +
\hat a_2^\dagger \hat a_2^\dagger \hat a_2 \hat a_2
+ \hat b_1^\dagger \hat b_1^\dagger \hat b_1 \hat b_1
+ \hat b_2^\dagger \hat b_2^\dagger \hat b_2 \hat b_2\right)
\cr &&
+ \ U_{ab} \left( \hat a_1^\dagger \hat a_1 \hat b_1^\dagger \hat b_1
+ \hat a_2^\dagger \hat a_2 \hat b_2^\dagger \hat b_2\right),
\label{oneH}
\end{eqnarray}
where we have assumed that intra-species interactions are the same in both gas components
and they are characterized by a coupling constant $U$.
The parameter $U_{ab}$ is a coupling constant that describes inter-species
interactions while $J$ stands for the tunneling rate between
the two potential wells. We assume also that numbers of particles of each component are
equal to $2N$. Such a choice of the system parameters allows us to perform fully analytical calculations. 
Analysis of a general case is beyond the scope of the present paper.
The Hamiltonian (\ref{oneH}) can be transformed to
\begin{eqnarray}
\label{h}
\hat H &=&- \ \frac{J}{2} \left(\hat a_1^\dagger\hat a_2
+\hat a_2^\dagger\hat a_1 + \hat b_1^\dagger\hat b_2
+\hat b_2^\dagger\hat b_1\right)
\cr &&
+\frac{U_s}{8} \left( \hat a_1^\dagger  \hat a_1
- \hat a_2^\dagger \hat a_2
+ \hat b_1^\dagger \hat b_1 - \hat b_2^\dagger \hat b_2\right)^2
\cr &&
+ \frac{U_d}{8} \left( \hat a_1^\dagger  \hat a_1 - \hat a_2^\dagger \hat a_2
- \hat b_1^\dagger \hat b_1 + \hat b_2^\dagger \hat b_2\right)^2,
\end{eqnarray}
where $U_s = U+U_{ab}$, $U_d = U-U_{ab}$ and constant terms have been omitted.
In the following we consider $U_s$ as the unit of energy.

\section{Perturbation approach}\label{pa}

\subsection{The second order effective Hamiltonian}\label{EH}

\begin{figure}
\centering
\includegraphics*[width=0.85\linewidth]{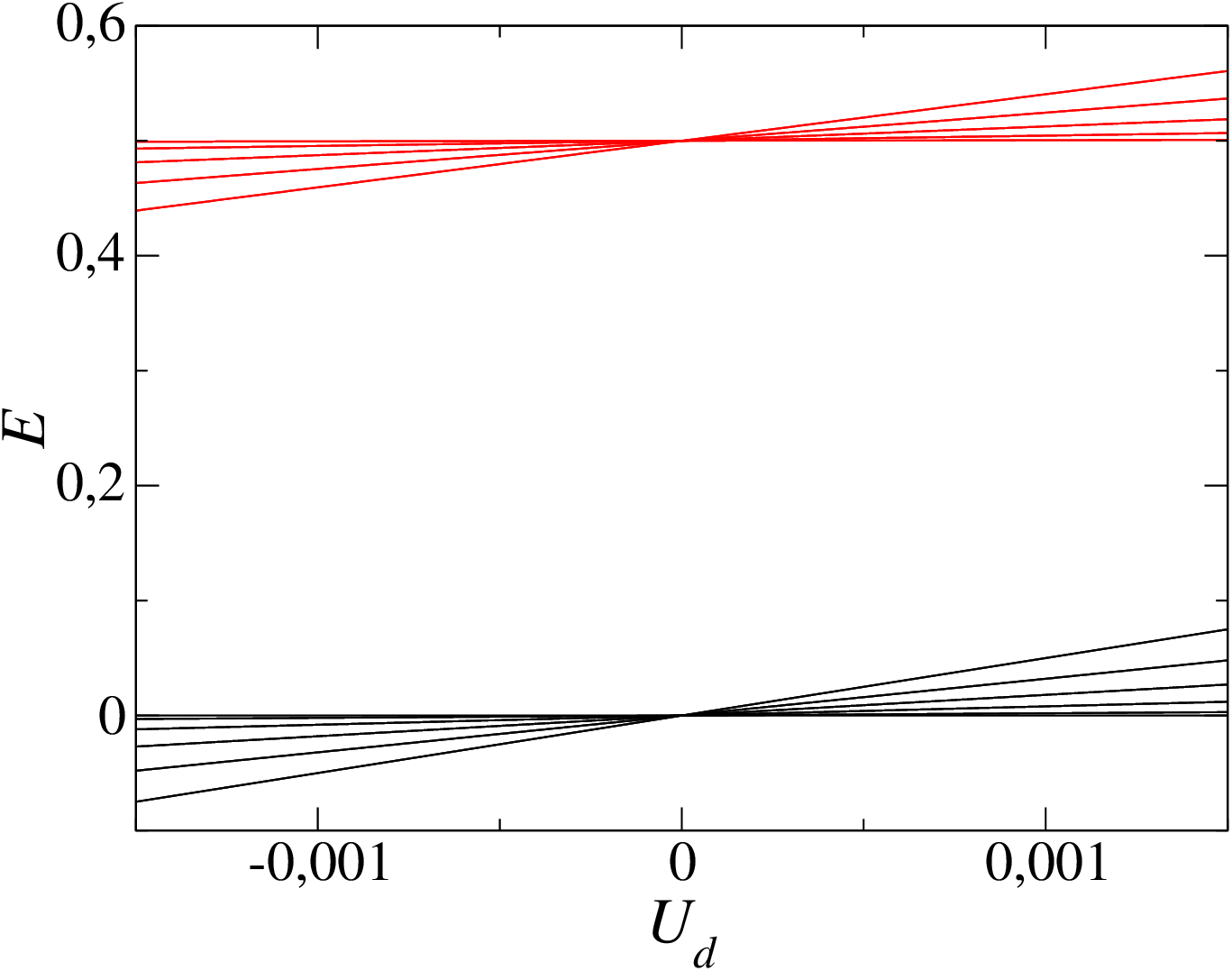}
\caption{(Color online) Energy levels (\ref{Eint2}) versus $U_d$ for a number of particles of
each component equal $2N=100$. Black lines: the lowest energy manifold, i.e.
$n_s=0$, red lines: the manifold corresponding to $|n_s|=1$.
}
\label{one}
\end{figure}

We are interested in the strong interaction limit. Therefore, the tunneling part of the Hamiltonian 
will be considered as a small perturbation. 
For $J=0$ the system Hamiltonian has exact eigenstates
\begin{equation}
|N + n_a,N-n_a \rangle  |N + n_b,N-n_b \rangle,
\label{fock}
\end{equation}
where  $N + n_a$, $N-n_a$ refer to numbers of particles of the component $a$ in the first and the second potential well, respectively,
and similarly for the component $b$.
The energies of such states (remember that $U_s$ is the unit of energy) are
\begin{equation}\label{Eint}
E= \frac{1}{2}(n_a+n_b)^2 + \frac{U_d}{2}(n_a-n_b)^2.
\end{equation}
Switching to variables $n_s = n_a+n_b$, $n_d = n_a-n_b$ we obtain eigenenergies
in a very simple form
\begin{equation}\label{Eint2}
E=\frac{1}{2}n_s^2 + \frac{U_d}{2}n_d^2.
\end{equation}
If we assume that the parameters satisfy the condition
\be \label{condition1}
1 \gg |U_d|N^2,
\ee
then manifolds with different values of $|n_s|$ are separated on the energy scale (see Fig. \ref{one}).
The lowest energy manifold is related to $n_s=0$ and states within each manifold are labelled by 
different values of $n_d$.  

Matrix elements of the tunneling part of the Hamiltonian are zero between states of the same manifold. 
However, this part of the Hamiltonian introduces couplings between different manifolds. 
In an effective Hamiltonian that describes the lowest manifold of the system the effect of the coupling 
can be included via the second order perturbation theory. 
A compact form of the effective Hamiltonian may be obtained if we introduce spin operators
\bea
\hat S_{jx}&=&\frac12(\hat a^\dagger_j\hat b_j+\hat b_j^\dagger \hat a_j), \cr
\hat S_{jy}&=&-\frac{i}{2}(\hat a^\dagger_j\hat b_j-\hat b_j^\dagger \hat a_j), \cr
\hat S_{jz}&=&\frac12(\hat a^\dagger_j\hat a_j-\hat b_j^\dagger \hat b_j). 
\eea
States belonging to the lowest manifold ($n_s=0$) can be written in the Fock basis (\ref{fock}) as 
\begin{equation} \label{state1}
|\psi \rangle = \sum_{n=-N}^N \psi(n)|N + n,N-n \rangle  |N - n,N+n \rangle.
\end{equation} 
The Fock states $|N + n,N-n \rangle|N - n,N+n \rangle$ are the eigenstates of the 
$\hat {\bf S}_j^2$, $\hat S_{1z}$ and
$\hat S_{2z}$ operators with the corresponding eigenvalues $N(N+1)$, 
$n$ and $-n$, respectively, so the $n_s=0$ manifold can be specified by:
\begin{equation}\label{condition}
\hat S_{1z} + \hat S_{2z}=0, \ \ \ \ \hat {\bf S}_j^2 = N(N+1).
\end{equation}
In the second order in $J$ the effective Hamiltonian that describes the lowest manifold reads 
\cite{kuklov03, altman03, isacsson05, svistunov, powell}
\begin{equation} \label{Ham1}
\hat H_2 =-2J^2\hat{\bf S}_1\cdot\hat{\bf S}_2+U_d(\hat S_{1z}^2+\hat S_{2z}^2). 
\end{equation}
The above Hamiltonian together with the condition (\ref{condition}) defines our problem
where each potential well is associated with an angular momentum operator 
and there is interaction between such subsystems due to tunneling of atoms.
Note that eigenstates of the Hamiltonian (\ref{Ham1}) depend on two parameters only, i.e. $U_d/(2J^2)$ and $N$.

\subsection{Classical limit}\label{CL}

Let us analyze the Hamiltonian (\ref{Ham1}) 
[in the manifold defined in (\ref{condition})] 
in the classical limit by substituting the spin operators by classical angular momentum components. The condition (\ref{condition}) implies that $S_{1z} = -S_{2z}$.
We are interested in the ground state of the system. The value of the tunneling part of the Hamiltonian
\be
-2J^2{\bf S}_1\cdot{\bf S}_2 = -2J^2(S_{1x}S_{2x} +S_{1y}S_{2y} + S_{1z}S_{2z} ),
\ee 
is minimal for $S_{1x}=S_{2x}$, $S_{1y}=S_{2y}$. 
For $U_d/(2J^2)+1 <0$ the ground state corresponds to $|S_{jz}|=N$ which can be related to 
$n_d = \pm 2N$ (i.e. phase separation occurs where different gas components occupy different potential wells).
For $U_d/(2J^2) +1 > 0$ the $z$-components of the angular momenta disappear in the ground state ($S_{jz}=0$) 
which corresponds to $n_d=0$ (i.e. equal mixture of both components in each potential well). 
When $U_d/(2J^2)+1=0$ we deal with the transition point where $|S_{jz}|$ can be arbitrary provided $S_{1z}+S_{2z}=0$.
Then all values of $n_d$ are equally probable (any mixture of both components in each potential well 
is equally likely). The transition between the phase separation and miscible regimes is discontinuous. 


\subsection{Quantum ground state}\label{QGS}

The analysis of the classical limit suggests that huge particle number fluctuations can be expected in the 
quantum ground state of the system at the transition point. 
Let us switch now to quantum analysis of the Hamiltonian (\ref{Ham1}). For $U_d=-2J^2$ the Hamiltonian 
reads 
\be \label{Ham2}
\hat H_2=J^2[(\hat S_{1x}-\hat S_{2x})^2+(\hat S_{1y}-\hat S_{2y})^2 - \hat {\bf S}_1^2 - \hat {\bf S}_2^2].
\ee
Applying the unitary (rotation) transformation
\begin{equation}
\hat S_{1x} \rightarrow - \hat S_{1x} \ \ \ \hat S_{1y} \rightarrow - \hat S_{1y} \ \ \ \hat S_{1z} \rightarrow  \hat S_{1z},
\end{equation}
which commutes with $\hat S_{jz}$ and thus leaves the manifold $\hat S_{1z} + \hat S_{2z}=0$ invariant, and
defining the total spin operator, $\hat {\bf S} = \hat {\bf S}_1 + \hat {\bf S}_2$,
we can rewrite the Hamiltonian (\ref{Ham2}) in the following form
\be \label{Ham3}
\hat H_2=J^2(\hat {\bf S}^2 -\hat {\bf S}_1^2 - \hat {\bf S}_2^2 ).
\ee 
States $|j,0 \rangle$ with an integer $j$, where $j(j+1)$ is an eigenvalue of the $\hat {\bf S}^2$ operator 
and $0$ is an eigenvalue of the $ \hat S_{z} = \hat S_{1z}+ \hat S_{2z}$ operator, are therefore eigenstates of our problem.
The energy spectrum reads
\begin{equation}\label{eigen}
E_{j} = J^2(j(j+1) - 2N(N+1)),
\end{equation}
with $0 \leq j \leq 2N$ and the ground state solution corresponds to $j=0$. 
In the basis (\ref{fock}) it takes the form of 
\be
\label{uniforms}
|\psi_0\ra=\frac{1}{\sqrt{2N+1}}\sum_{n=-N}^N|N + n,N-n \rangle  |N - n,N+n \rangle.
\ee
Huge particle number fluctuations become apparent in
Eq.~(\ref{uniforms}) where the ground state turns out to be a uniform superposition of all Fock states 
belonging to the lowest energy manifold ($n_s=0$). That is, all values of $n_d$ are equally probable.

\begin{figure}
\centering
\includegraphics*[width=0.9\linewidth]{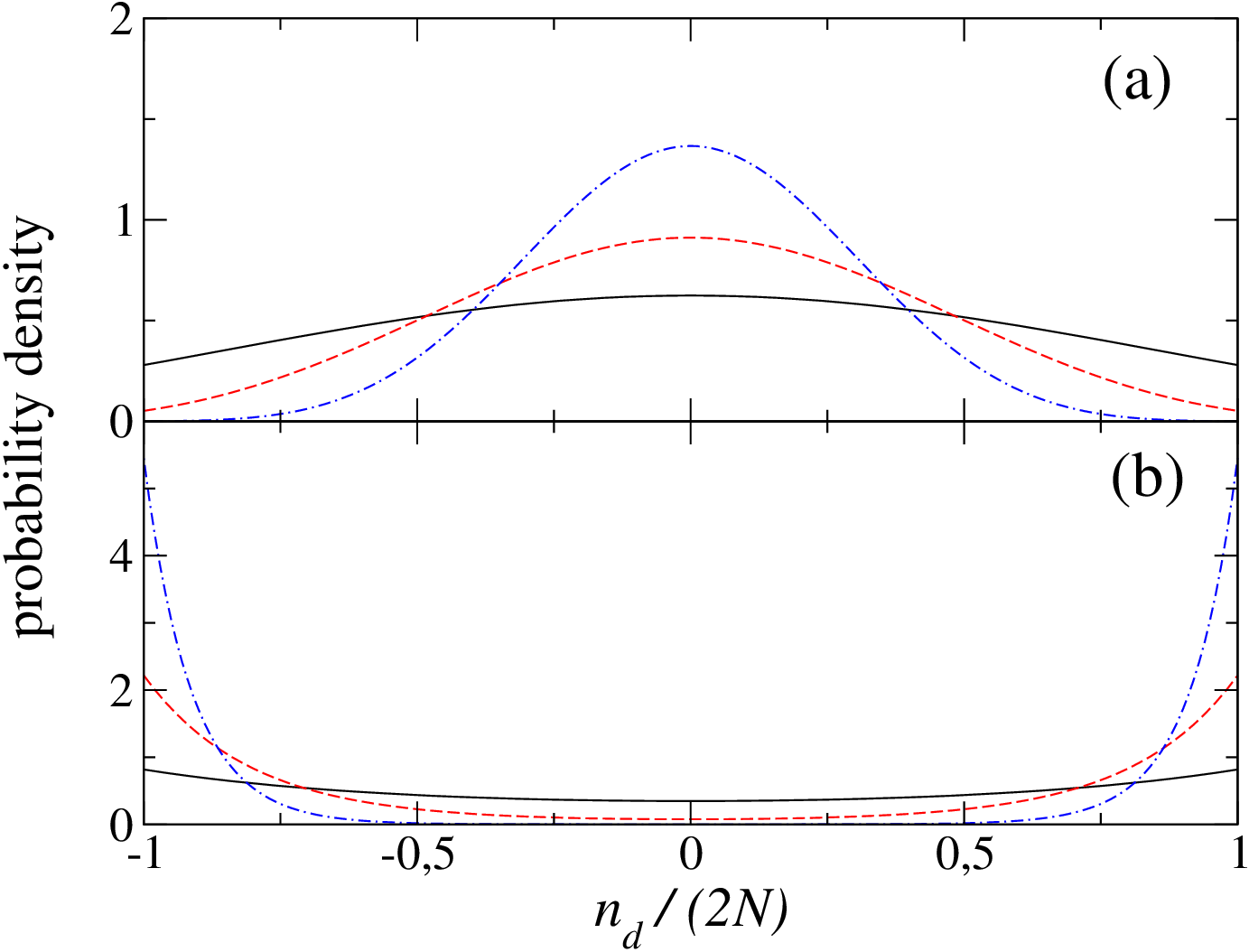}
\caption{(Color online) Ground state probability density   $\rho(n_d)=|\psi_0(n_d)|^2$
for $\frac{U_d}{2J^2}+1= 10^{-3}$ (a) and $\frac{U_d}{2J^2}+1=- 10^{-3}$ (b). Solid black lines are related to
the number of particles in each gas component $2N=50$, dashed red lines to $2N=100$ and dotted-dashed blue lines to
$2N=200$.
}
\label{two}
\end{figure}

It is interesting to note that we can construct the exact quantum ground state using the
superposition of symmetry broken solutions obtained in the classical limit.
At the transition point the classical analysis tells us that in the ground state the potential wells are 
associated with classical angular momenta where orientation of one is given by 
$(\theta_1,\phi_1)=(\theta,\phi)$ and the other one by $(\theta_2,\phi_2)=(\pi-\theta,\phi)$ 
and spherical angles $\theta$ and $\phi$ can be arbitrary. 
The best quantum approximation of a classical angular momentum is a coherent state \cite{thomas, glauber}
\bea
|\theta_j,\phi_j\ra&=&\frac{1}{\sqrt{(2N)!}} \left( \cos \frac{\theta_j}{2} e^{i\phi_j/2} \hat a_j^\dagger \right. \cr
&& \left. + \sin \frac{\theta_j}{2} 
e^{-i\phi_j/2} \hat b_j^\dagger   \right)^{2N} |0\rangle.
\label{cohst0}
\eea
If we postulate that the quantum ground state of our system can be approximated by a single tensor product state
$|\theta,\phi\ra|\pi-\theta,\phi\ra$ the rotational symmetry of the Hamiltonian (\ref{Ham1}) will be broken.
A rotationally invariant state can be restored if we prepare a uniform superposition of the tensor product states, i.e. 
by integrating over all solid angles
\be
|\psi_0\ra=\frac{\sqrt{2N+1}}{2}\int_0^{2\pi} \frac{\mbox{d}\phi}{2\pi} \int_0^\pi \sin \theta \mbox{d} \theta \, 
|\theta,\phi\rangle |\pi - \theta, \phi \rangle,
\label{cohst}
\ee
where $\la\psi_0|\psi_0\ra=1$.
Substituting Eq.~(\ref{cohst0}) into Eq.~(\ref{cohst}), integrating over $\phi$ and employing the identity
\bea
\int_{-1}^1 \mbox{d}(\cos\theta) \left(\frac{1+\cos\theta}{2} \right)^{N+n}
\left(\frac{1-\cos\theta}{2} \right)^{N-n} = 
\cr 
\frac{2}{2N+1}\left[\binom{2N}{N+n}\right]^{-1} ,   
\eea
that follows from the completeness relation of the coherent states we restore the quantum ground state (\ref{uniforms}).

We see that at the transition point the classical analysis allows us to construct the exact ground state of the system. However, in 
the close vicinity of the transition point this analysis is not able to provide a good estimate for the ground state of the system.
This is because the classical approach predicts discontinuous transition between the phase separation and miscible regimes
but the transition is actually continuous. Starting with the classical ground states and using the coherent states we can
construct quantum ground states that depend on $N$ and on the sign of the parameter $U_d/(2J^2)+1$ but not on its absolute value. 
However, the exact diagonalization indicates that there is a range of $U_d/(2J^2)+1$ where the ground state changes continuously 
from the miscible to phase separation character. As one can expect this range shrinks with $N$ because
differences between the classical and quantum angular momentum diminish in the large $N$ limit. 
This is illustrated in Fig.~\ref{two} where we plot $\rho(n_d)=|\psi_0(n_d)|^2$ 
for $U_d/(2J^2)+1=10^{-3}$ and $-10^{-3}$, i.e. close to the transition point, for different 
values of $N$ obtained in numerical diagonalization of the effective Hamiltonian (\ref{Ham1}).

\begin{figure}
\centering
\includegraphics*[width=0.9\linewidth]{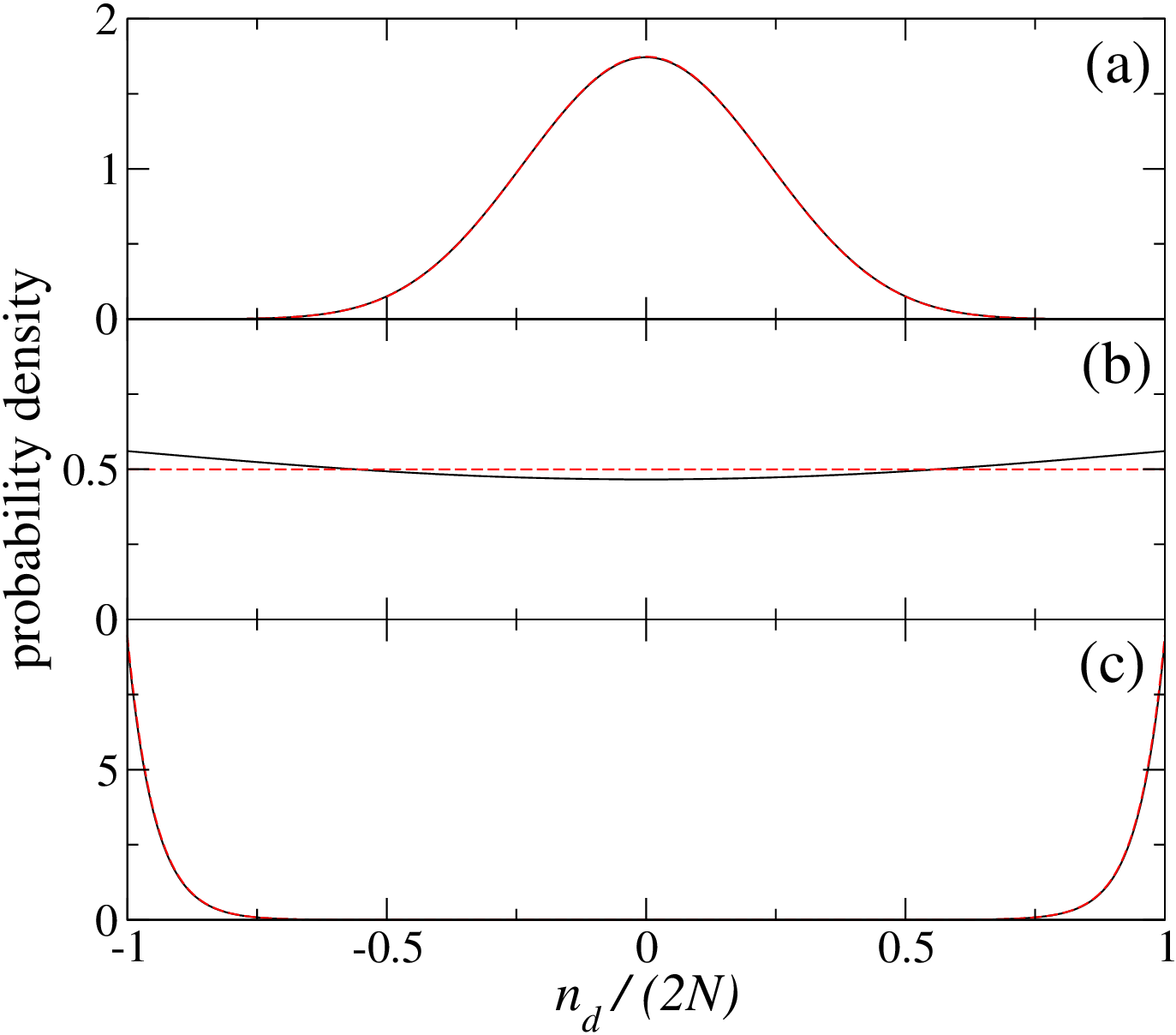}
\caption{(Color online) Ground state probability densities  for a number
of particles in each component $2N=1000$, $J=5\times10^{-7}$ and $\frac{U_d}{2J^2}= - 0.9999$ (a),
$\frac{U_d}{2J^2}=-1$ (b) and $\frac{U_d}{2J^2}=-1.0001$ (c). Solid black lines correspond to
$\rho(n_d)=\sum_{n_s}|\psi_0(n_s,n_d)|^2$ where $\psi_0(n_s,n_d)$ is obtained by diagonalization of (\ref{h}) while dashed red
lines are related to $\rho(n_d)=|\psi_0(n_d)|^2$ with $\psi_0(n_d)$ obtained by diagonalization of the effective Hamiltonian (\ref{Ham1}). In panels (a) and (c) 
the black and red lines are hardly distinguishable.
}
\label{three}
\end{figure}

\subsection{Validity of the perturbation approach}\label{V}

Our predictions are based on the effective Hamiltonian (\ref{Ham1}) which is second order in $J$, and they are correct provided higher order terms can be neglected. 
The fourth order terms are smaller than $J^4(2N)^4$ or $J^2|U_d| (2N)^4$.
At the transition point $|U_d|=2J^2$, so if
$J^4(2N)^4$ is much smaller than the energy gap between the ground and first excited states of the Hamiltonian 
(\ref{Ham3}), i.e. $E_1-E_0 = 2J^2$, then the higher order terms can be neglected and the system is properly described 
by the second order Hamiltonian (\ref{Ham3}). Hence, 
\be
J^2 (2N)^4 \ll 1,
\label{condition3}
\ee
is a sufficient condition for the validity of the ground state (\ref{uniforms}) that describes
the huge particle number fluctuations in the system.

We have tested our predictions comparing them with exact numerical calculations. Figure \ref{three} shows probability densities
$\rho(n_d)$ corresponding to ground states of the system obtained by diagonalization of the full Hamiltonian (\ref{h}) and the effective Hamiltonian (\ref{Ham1}) for $2N=1000$ and $J= 5\times10^{-7}$.
Different panels are related to different values of $U_d$ in the vicinity of the transition point. That is, $\frac{U_d}{2J^2}+1= 10^{-4}$ corresponds to the miscible regime, $\frac{U_d}{2J^2}+1=0$ to the transition point and $\frac{U_d}{2J^2}+1= - 10^{-4}$ to the phase separation side of the transition point. We can see that on both sides of the transition point the density $\rho(n_d)$ is peaked around the classical solutions, while at the transition point it is uniformly distributed. Figure~\ref{three} indicates perfect agreement between the perturbation calculations and the exact results even though $J^2(2N)^4 = 0.25$ and thus 
the condition (\ref{condition3}) is barely fulfilled.

\section{Conclusion}\label{Con}

In summary, we have analyzed a strongly interacting two-component Bose gas in a double well potential for parameters close to the transition point where the phase separation occurs.  
The second order effective Hamiltonian allows us to describe 
the system in the vicinity of the transition point when higher order corrections are negligible. We have shown that, at the
transition point, the ground state of the system becomes a uniform superposition of Fock states. That is, the system 
reveals huge quantum fluctuations of populations of the potential wells.

\section*{Acknowledgment}

We are grateful to Maciej Lewenstein for a fruitful discussion and his help in solving the Hamiltonian 
eigenvalue problem at the transition point.
Support within Polish Government scientific funds (for years 2008-2011 -- PZ and
KS, 2009-2012 -- BO) as a research project is acknowledged.


\begin{thebibliography}{99}


\bibitem{leggett01} A. J. Leggett, Rev. Mod. Phys. {\bf 73}, 307 (2001).

\bibitem{ladd10} T. D. Ladd, F. Jelezko, R. Laflamme, Y. Nakamura, C. Monroe,
and J. L. O'Brien, Nature {\bf 464}, 45 (2010).

\bibitem{dunningham04} J. A. Dunningham and K. Burnett, Phys. Rev. A {\bf 70}, 033601 (2004).

\bibitem{oberthaler07} R. Gati and M. K. Oberthaler, J. Phys. B {\bf 40}, R61 (2007).

\bibitem{bloch08} I. Bloch, J. Dalibard, and W. Zwerger, Rev. Mod. Phys. 80, 885 (2008).

\bibitem{esteve08} J. Esteve, C. Gross, A. Weller, S. Giovanazzi, and M. K. Oberthaler,
	Nature 455, 1216 (2008).

\bibitem{orzel01} C. Orzel, A. K. Tuchman, M. L. Fenselau, M. Yasuda, and M. K. Kasevich, Science 291, 2386 (2001).

\bibitem{choi05} S. Choi and N. P. Bigelow, Phys. Rev. A 72, 033612 (2005).

\bibitem{SFMI} D. Jaksch, C. Bruder, J. I. Cirac, C. W. Gardiner, and
	P. Zoller, Phys. Rev. Lett. 81, 3108 (1998).

\bibitem{greiner02} M. Greiner, O. Mandel, T. Esslinger, T. W. H\"ansch, and
	I. Bloch, Nature 415, 39 (2002).

\bibitem{gerbier06} F. Gerbier, S. Foelling, A. Widera, O. Mandel, and I. Bloch
	Phys. Rev. Lett. 96, 090401 (2006).

\bibitem{break} J.I. Cirac, M. Lewenstein, K. Molmer, and P. Zoller, Phys. Rev. A 57, 1208 (1998).

\bibitem{walls98} J. Ruostekoski, M.J. Collett, R. Graham, and Dan. F. Walls, Phys. Rev. A 57, 511 (1998).

\bibitem{gordon99} D. Gordon and C. M. Savage , Phys. Rev. A 59, 4623 (1999).

\bibitem{castin01} Y. Castin, in {\it Les Houches Session LXXII, Coherent
atomic matter waves 1999}, edited by R. Kaiser, C. Westbrook
and F. David, (Springer-Verlag Berlin Heilderberg
New York 2001).

\bibitem{dziarmaga10} J. Dziarmaga, P. Deuar, and K. Sacha, Phys. Rev. Lett. {\bf 105}, 018903 (2010).

\bibitem{miszmasz} R. V. Mishmash, and L. D. Carr, Phys. Rev. Lett. {\bf 105}, 018904 (2010).

\bibitem{ho04} T.-L. Ho and C. V. Ciobanu, J. Low Temp. Phys. 135, 257 (2004).

\bibitem{jack05} M. W. Jack and M. Yamashita, Phys. Rev. A 71, 023610 (2005).





\bibitem{kuklov03} A. B. Kuklov and B. V. Svistunov, Phys. Rev. Lett. {\bf 90}, 100401 (2003).

\bibitem{altman03} E. Altman, W. Hofstetter, E. Demler, M. D. Lukin, New J. Phys. {\bf 5}, 113 (2003).

\bibitem{isacsson05} A. Isacsson, M.-C. Cha, K. Sengupta, and S. M. Girvin, Phys. Rev. B {\bf 72}, 184507 (2005).

\bibitem{svistunov} S. G. S\"oyler, B. Capogrosso-Sansone, N. V. Prokof'ev, and B. V. Svistunov, arXiv:0811.0397.

\bibitem{powell} S. Powell, arXiv:0902.1993.


\bibitem{thomas} F.T. Arecchi, E. Courtens, G. Gilmore, H. Thomas, Phys. Rev. A {\bf 6}, 2211 (1972).

\bibitem{glauber} R. Glauber, F. Haake, Phys. Rev. A {\bf 13}, 357 (1976).


\end{thebibliography}
\end{document}